\documentclass[fleqn,10pt]{wlscirep}
\usepackage[utf8]{inputenc}
\usepackage{tikz}
\usepackage{pgfplots}
\pgfplotsset{compat=1.17}
\usepackage[T1]{fontenc}
\title{Towards Full Integration of Artificial Intelligence in Colon Capsule Endoscopy's Pathway}

\author[1,*]{Esmaeil S. Nadimi}
\author[1]{Jan-Matthias Braun}
\author[2,3]{Benedicte Schelde-Olesen}
\author[1]{Emile Prudhomme}
\author[1]{Victoria Blanes-Vidal}
\author[2,3]{Gunnar Baatrup}

\affil[1]{Applied AI and Data Science (AID), Maersk Mc-Kinney Moller Institute, Faculty of Engineering, University of Southern Denmark, Odense, Denmark}
\affil[2]{Department of Surgery, Odense University Hospital, Odense, Denmark}
\affil[3]{Department of Clinical Research, University of Southern Denmark, Odense, Denmark}

\affil[*]{esi@mmmi.sdu.dk}

\begin{abstract}
Despite recent surge of interest in deploying colon capsule endoscopy (CCE) for early diagnosis of colorectal diseases, there
remains a large gap between the current state of CCE in clinical practice, and the state of its counterpart optical colonoscopy (OC). Our study is aimed at closing this gap, by focusing on the full integration of AI in CCE's pathway, where image processing steps linked to the detection, localization and characterisation of important findings are carried out autonomously using various AI algorithms. We developed a recognition network, that with an impressive sensitivity of $99.9\%$, a specificity of $99.4\%$, and a negative predictive value (NPV) of $99.8\%$, detected colorectal polyps. After recognising a polyp within a sequence of images, only those images containing polyps were fed into two parallel independent networks for characterisation, and estimation of the size of those important findings. The characterisation network reached a sensitivity of $82\%$ and a specificity of $80\%$ in classifying polyps to two groups, namely neoplastic vs. non-neoplastic. The size estimation network reached an accuracy of $88\%$ in correctly segmenting the polyps. By automatically incorporating this crucial information into CCE's pathway, we moved a step closer towards the full integration of AI in CCE's routine clinical practice.  
\end{abstract}
\begin{document}

\flushbottom
\maketitle
%
%
\thispagestyle{empty}

\section*{Introduction}

Colon capsule endoscopy (CCE) is a non-invasive procedure offering advantages in diagnosis, follow up, and management of colorectal diseases, over its counterparts optical colonoscopy (OC), flexible sigmoidoscopy and computed tomographic colonography (CTC) \cite{Gunnar,CCE,VICOCA}. Based on several clinical trials, we know that CCE is superior to CTC in the detection of polyps larger than 6mm and non-inferior for those larger than $10 mm$. We also know that in patients with incomplete OC, CCE shows a higher diagnostic yield compared to CTC for any size polyp. Additionally, patient's preference to undergo CCE rather than OC, and a significantly lower complication rate of CCE compared to OC justifies the goal of achieving a more widespread use of CCE in clinical practice. However, a few important shortcomings have hindered reaching a unanimous support of CCE, among medical professionals \cite{koulaouzidis2023current}. First and foremost, bowel cleansing quality plays a major role in the diagnostic yield of CCE \cite{Buijs2018Cleanliness,Benedicte2023}. Besides, logistical challenges around the delivery and collection of the capsule and equipment, tedious manual assessment  of roughly $12,000$ retrieved images from each patient investigation to look for important findings, low resolution white-light imaging, and low completion rates are some additional challenges in routine clinical practice. To address most of these shortcomings, we \cite{sahafi2022edge} designed an artificial intelligence (AI)-empowered wireless video endoscopic capsule that featured 1) real-time on-the-go image processing, 2) enhanced visualization of the mucous layer using both white-light (WLI) and narrow-band imaging (NBI), and 3) bi-directional communication with patient’s personal electronic devices to report important findings. The design and development of this novel edge-AI solution significantly advanced the state of current endoscopic capsules within both hardware and software, optimizing the diagnostic pathway of CCE by improving image quality, and more importantly, by paving the way towards the integration of AI in real-time analysis of retrieved images from the investigations. Despite reaching these significant milestones, there remains a big gap between the current state of this solution and its clinical implementation, that raise concerns among medical professionals about the feasibility of this solution in clinical practice.  

The pathway for CCE involves several stages, including patient's bowel preparation, capsule ingestion followed by manual image analysis, post-procedure care and treatment. In this study, the main goal is to focus on full integration of AI in the image analysis stage, in which detection, localization and characterisation of important findings are carried out autonomously using various AI algorithms. It is worth mentioning that by characterisation, we refer to the analysis of abnormality's morphology in terms of estimated size, and histopathology analysis results (HP) in terms of neoplastic properties. Building upon our previous studies on the detection and localisation of colorectal polyps, \cite{Benedicte2023,Benedicte2023Landmarks,blanes2018matching,Buijs2018Cleanliness,Herp,NADIMI1,Tashk2022AIDUNet,Tashk2022DetectandSegment,Ulrik,Victoria1}, and by incorporating crucial information on both estimated size and histopathology (HP) of important findings, we optimize CCE's pathway and move a step closer towards the full integration of AI in CCE's routine clinical practice. 

The literature on AI-based detection and classification of colorectal polyps, retrieved from optical colonoscopy, contains a relatively rich body of work \cite{LOU2023102341}. However, excluding our publications, only a handful of studies (\cite{moen2022artificial,saraiva2021artificial,yamada2021automatic}) focused on similar research questions within CCE investigations. This is due to the fact that unlike optical colonoscopy where databases such as Kvasir (A Multi-Class Image Dataset for Computer Aided Gastrointestinal Disease Detection) are publicly available, a very limited resources of data on CCE investigations exist. Besides, CCE's invariant imaging modality (WLI) has contributed to a significantly lower-resolution images, challenging the success of any AI algorithm that can live up to the standards required for a robust solution in routine clinical practice. 

This study is built upon a recently completed "Danish CareForColon2015 trial (cfc2015)", launched in 2021 as part of the Danish Colorectal Cancer Screening program, being the largest randomized controlled trial to date on CCE \cite{GovTrial}. The primary aim of the trial was to compare the number of detected colorectal cancers, and intermediate- and high-risk adenomas between the intervention and control groups. In addition, the cfc2015 trial had a number of secondary aims, including estimating patient acceptability,
complication rate, completion rate, interval CRC rate, patient reported outcomes (PRO), long-term cancer incidence rate, social inequality, improvement of CCE applicability, and cost-effectiveness of the trial structure of the active group. Out of the $370,306$ invited individuals to screen for CRC and after adjustment for assumptions on dropout rate, preference proportion, fecal immunochemical test (FIT) positive rate, and FIT participation rate, a total of $2015$ FIT-positive patients underwent CCE investigations. CCE videos retrieved from these $2015$ patient investigations formed the basis for the development and validation of the AI algorithms that are presented in this study. For more details on cfc2015 trial, we refer interested readers to visit \href{https://clinicaltrials.gov/ct2/show/NCT04049357}{https://clinicaltrials.gov/ct2/show/NCT04049357} or \cite{Ulrik}. 

\begin{figure}[ht]
\centering
\includegraphics[width=\linewidth]{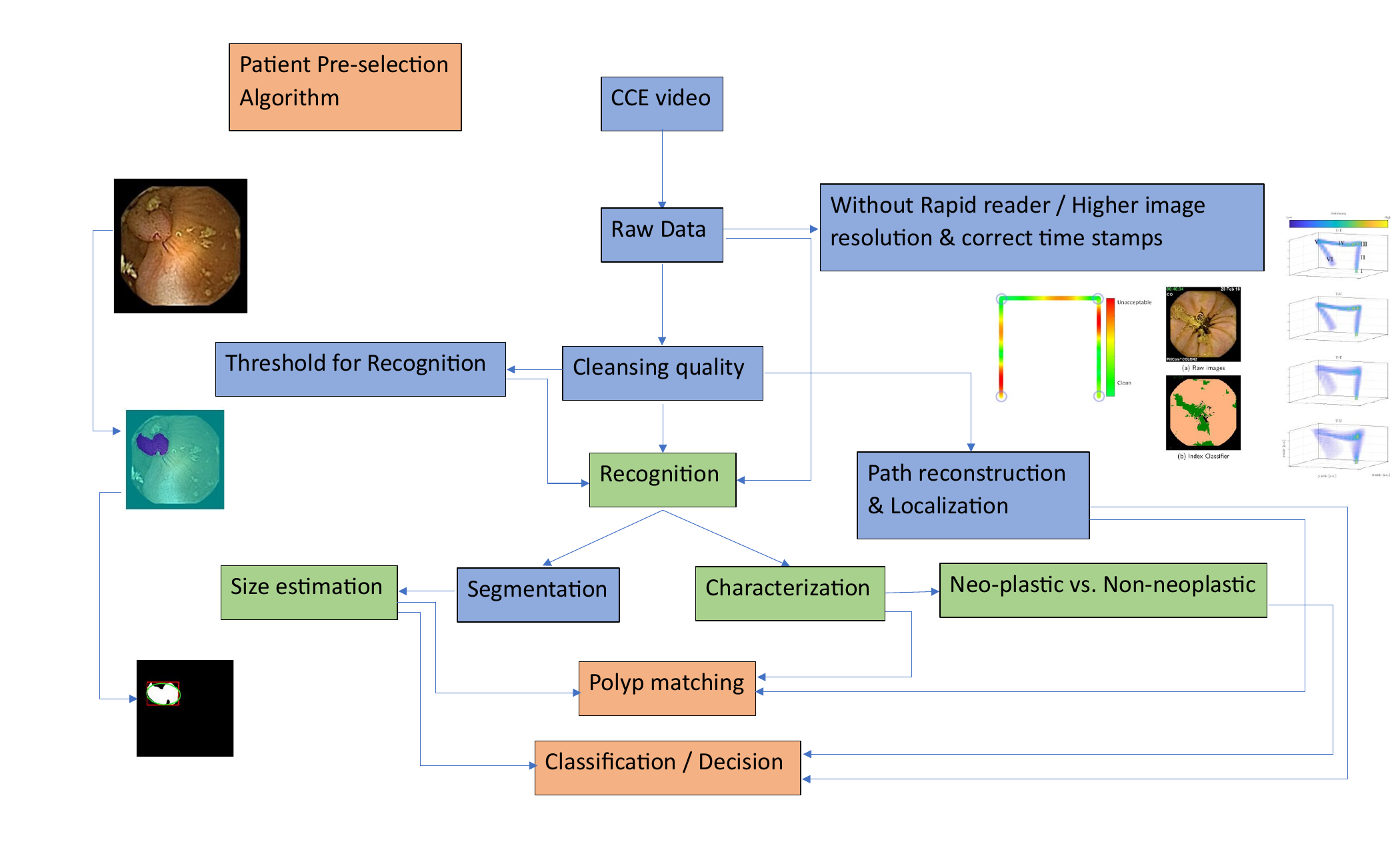}
\caption{AI-based optimization of CCE's pathway (image analysis).}
\label{fig1}
\end{figure}

\begin{figure}[ht]
\centering
\includegraphics[width=\linewidth]{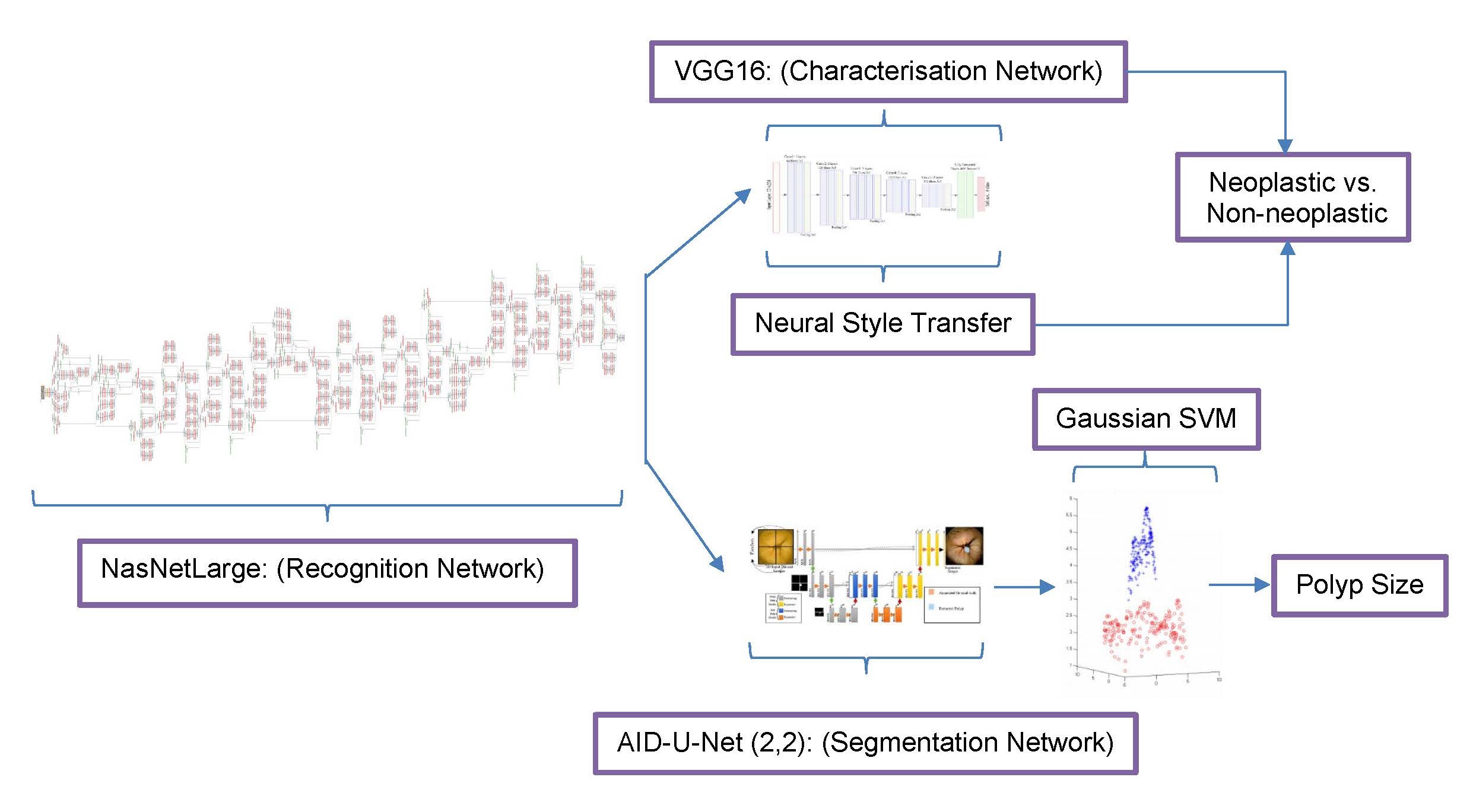}
\caption{Sequence of neural networks (NN) for recognition, characterization and estimation of size of polyps.}
\label{fig_method}
\end{figure}

\subsection*{Contribution}

The sketch of the workflow associated with CCE's pathway automation is presented in Figure \ref{fig1} and Figure \ref{fig_method}. The steps presented in blue have been published in previous works\cite{Buijs2018Cleanliness,Benedicte2023,Herp,Tashk2022AIDUNet,Tashk2022DetectandSegment}, while those in orange are under development. This paper presents those steps highlighted in green, which focus on the recognition of important abnormalities, estimating their size and define their histopathology. 

The organization of this paper is as follows: We first present a deep neural network, that is able to detect abnormalities with an impressive sensitivity and specificity. Images on these found abnormalities are then fed into two separate parallel algorithms for size estimation and characterization. Details of these algorithms are provided in the article. We then discuss the outcomes of these three algorithms, present their strengths and shortcomings, and conclude on the feasibility of optimizing CCE's image analysis pathway using AI.  

\section*{Methods and Results}

\subsection*{Ethics}
The study was approved by the Regional Health Research Ethics committee (journal number S-20190100), was registered with the Regional Data Protection Agency (journal number 19/29858), as well as with ClinicalTrials (identifier NCT04049357). All participants received verbal and written study information prior to participation and signed informed consent was obtained from each individual. The study was conducted in accordance with the declaration of Helsinki.

\subsection*{Recognition}

Our previous research on the detection and localization of colorectal polyps in a dataset of approximately $800$ images ($400$ featuring polyps and $400$ of normal mucosa) using an improved version of ZF-Net resulted in an accuracy of $98.0\%$, a sensitivity of $98.1\%$ and a specificity of $96.3\%$ \cite{NADIMI1}. Despite the fact that our improved ZF-Net still remains as one of the best performing networks in the literature, having the prevalence of CRC among the FIT positive population in mind implies that, by deploying this network for the cfc2015 trial, we accept the risk of missing $4$ potential cancers within the FIT positive population consisting of $2015$ patients. Besides, the findings from this DNN form the basis for the input variables to be fed into both size estimation and characterization algorithms. Keeping these two conditions in mind calls for a new DNN with a higher negative-predictive-value (NPV), and therefore an elevated sensitivity and specificity. After testing almost all the main architectures in the literature such as ResNet50, InceptionV3, etc., we deployed NasNetLarge as the backbone architecture for the recognition of abnormalities.     

\begin{figure}[ht]
\centering
\includegraphics[scale=0.1]{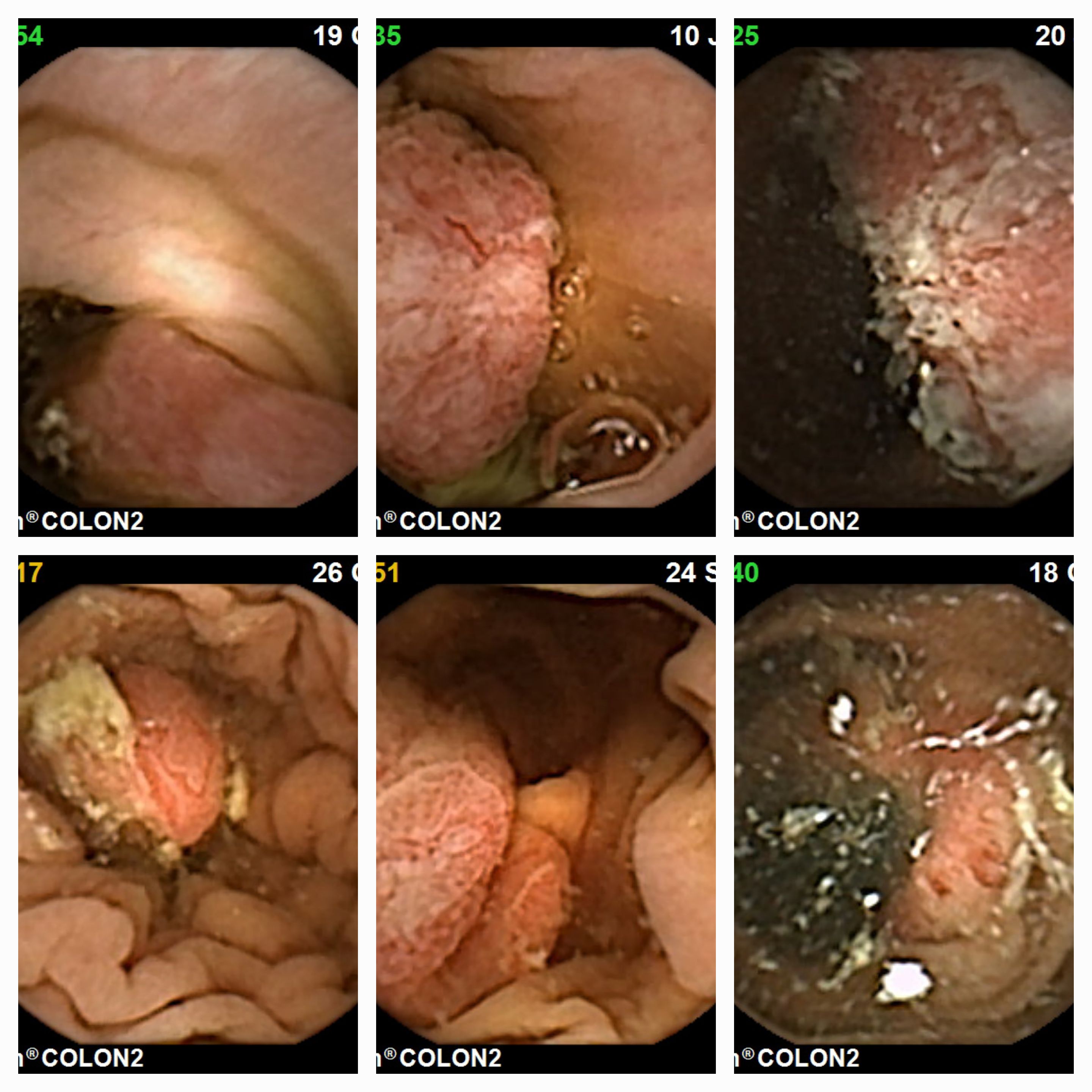}
\caption{Training for abnormality detection.}
\label{fig2}
\end{figure}

To adapt network's architecture and to use transfer learning for the purpose of this study, we modified the last $20$ learnable layers, and froze the parameters of the remaining layers, accordingly. The learning rate was initially set to $1e-3$, but adaptively reduced after every $2$ epochs during the training process, until the validation criteria were met. The epoch size for training process was limited to a maximum of $6$, with a mini batch size of $10$, and the validation frequency of $798$. The dataset containing images of both colorectal polyps ($1751$) and the normal mucosa ($1672$) (Figure \ref{fig2}) was augmented by horizontal and vertical random reflection, random scaling and random translation, along with random rotation, all picked from a continuous uniform distribution, without affecting the contents or the size of the images. This augmentation resulted in a database containing $5838$ images of colorectal polyps and $5573$ images of normal mucous layer. The dataset was split to $70\%$ for both training and validation process, while the remaining $30\%$ was allocated to the test process. NasNetLarge with the aforementioned configuration resulted in a sensitivity of $99.9\%$, a specificity of $99.4\%$, and an NPV of $99.8\%$ on the test set, implying that $<1$ cancers among the cfc2015 trial subjects will be missed.

\begin{figure}[ht]
\centering
\includegraphics[scale=0.1]{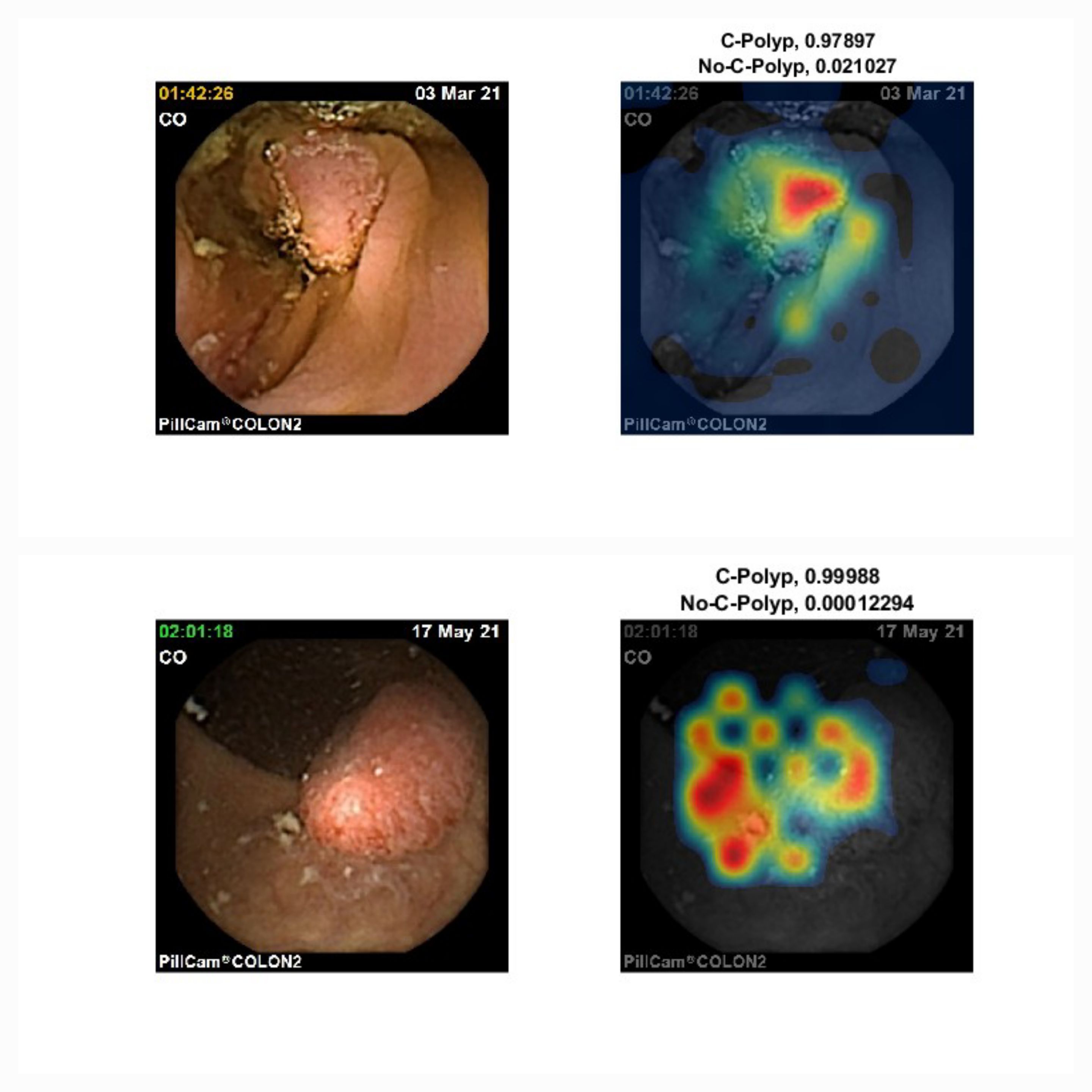}
\caption{Heat map, and network's explanation on decision making process. C-Polyp and No-C-Polyp stand for colorectal polyp and no colorectal polyp, respectively.}
\label{fig:stream}
\end{figure}

To evaluate as whether the network has learnt sufficient features, and knowledge has been transferred, we exposed the recognition network to an additional set of images as presented in Figure \ref{fig:stream}. Manual examination performed by trained CCE readers had classified $3$ specific cases as either inflammation or normal tissue. These diagnoses were confirmed by the specialists. When the medical experts in charge of cfc2015 trial analyzed the images as part of a routine quality check, they realized these images were misdiagnosed, and identified these three cases as cancers. To test the performance of our DNN, the images retrieved from those $3$ missed cancers were fed into the recognition network as part of the test set, and were all correctly classified as important findings, confirming generalization capabilities of our network. It is worth mentioning that due to this incidence, the medical experts in charge of cfc2015 trial reevaluated all the images retrieved from the patients, ensuring that there was no important finding being overlooked.  

\subsection*{Size estimation}

The size estimation algorithm is built upon our previous work on the development of a novel semantic segmentation network called AID-U-Net \cite{Tashk2022AIDUNet,Tashk2022DetectandSegment}. AID-U-Net consists of direct contracting and expansive paths, along with a distinguishing feature of sub-contracting and sub-expansive paths. This architecture resulted in AID-U-Net outperforming all the state-of-the-art solutions with an $F_1$ score of $88.1\%$ compared to networks such as U-Net ($81.1\%$) and U-Net++ ($87.6\%$), where no specific pre-trained backbones were needed. We further showed that the best performing AID-U-Net for segmenting CCE images featured a depth of two for both direct path and sub-path (AID-U-Net(2,2)). For more details on the architecture of AID-U-Net and its performance, we refer interested readers to \cite{Tashk2022AIDUNet}.  

\begin{figure}[ht]
\centering
\includegraphics[scale=0.05]{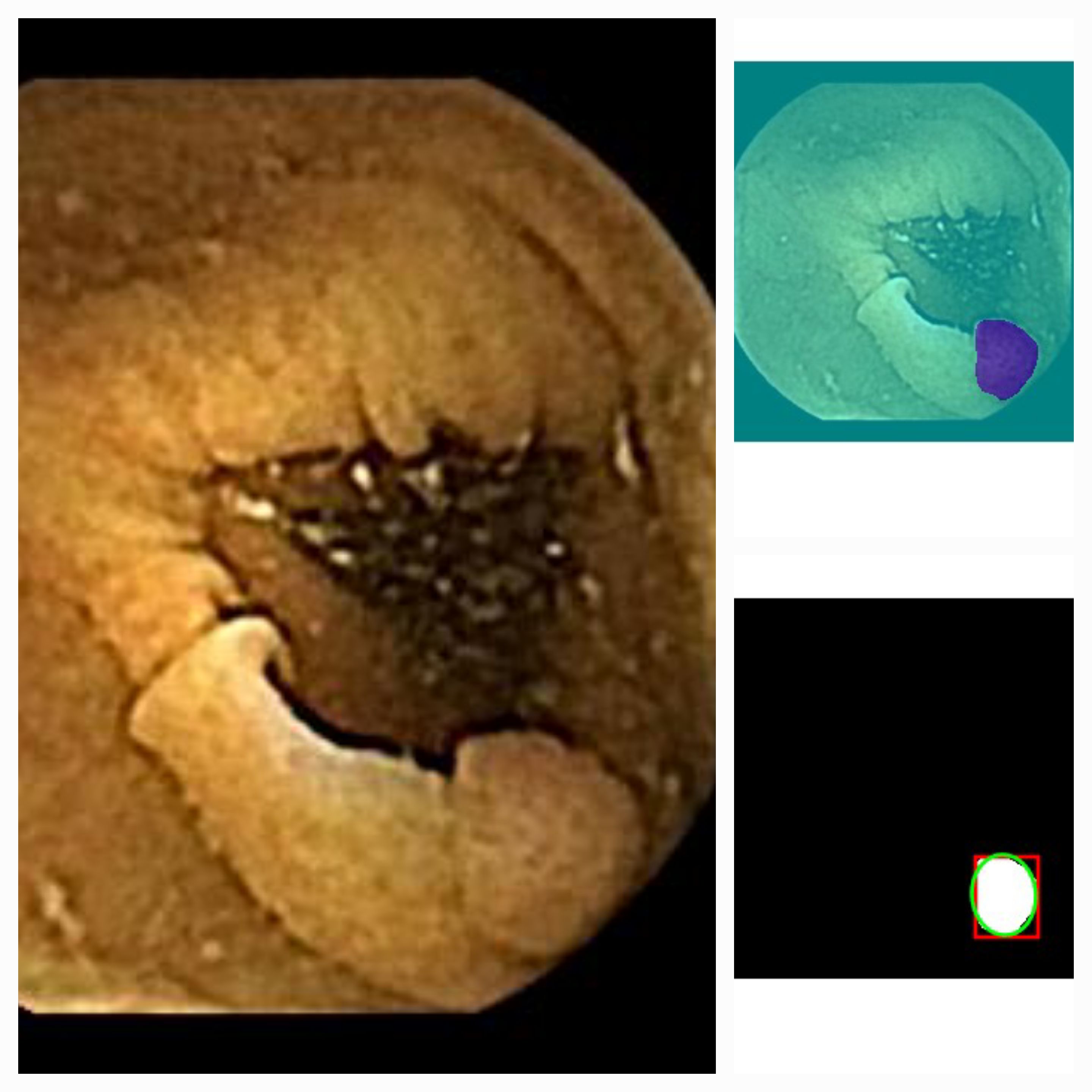}
\includegraphics[scale=0.05]{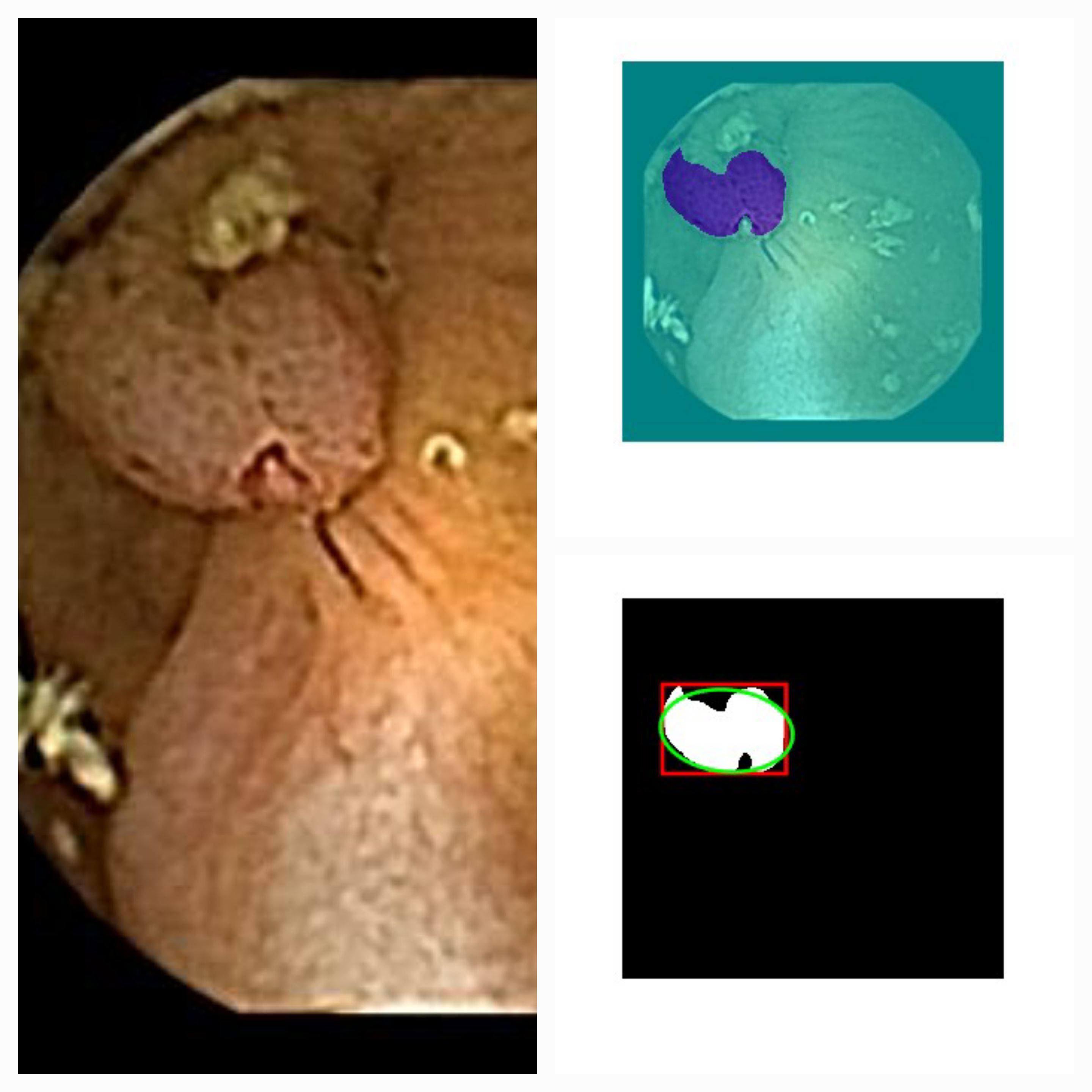}
\caption{Segmentation and size estimation.}
\label{fig:AID-U-Net1}
\end{figure}

We applied AID-U-Net(2,2) to the dataset of augmented images containing colorectal polyps ($5838$), and a total of $4685$ ($81\%$) images were correctly segmented. Implementing the same performance metric for U-Net and U-Net++ resulted in $61\%$ and $72\%$ correct segmentation, respectively. In this study, incorrect segmentation refers to three scenarios: 1) missing a ROI (polyp), 2) segmenting the wrong region, or 3) splitting one ROI e.g. a large polyp into two or more segments. It is worth mentioning that, by assuming that each image only contains one ROI, and by adding the estimated sizes of all the regions segmented by AID-U-Net(2,2) together, the accuracy of segmentation task was improved to $88\%$. A few examples of segmented polyps along their corresponding bounding boxes and fitted ellipses are presented in Figure \ref{fig:AID-U-Net1}. 

A precise estimation of the size requires information of the depth of a polyp with respect to the camera lens, which in CCE, is not available. As a result, our size estimation algorithm simply measures the ratio between the largest diameter of the fitted ellipse to the segmented polyp, and the total size of the image, where the additional peripheral information (date and time) are cropped out. Following this strategy, we reached a perfect match between our estimated size and the outcome from the RAPID Reader \cite{Rapid}, used by trained CCE readers.

\begin{figure}[ht]
\centering
\includegraphics[width=10cm,height=10cm]{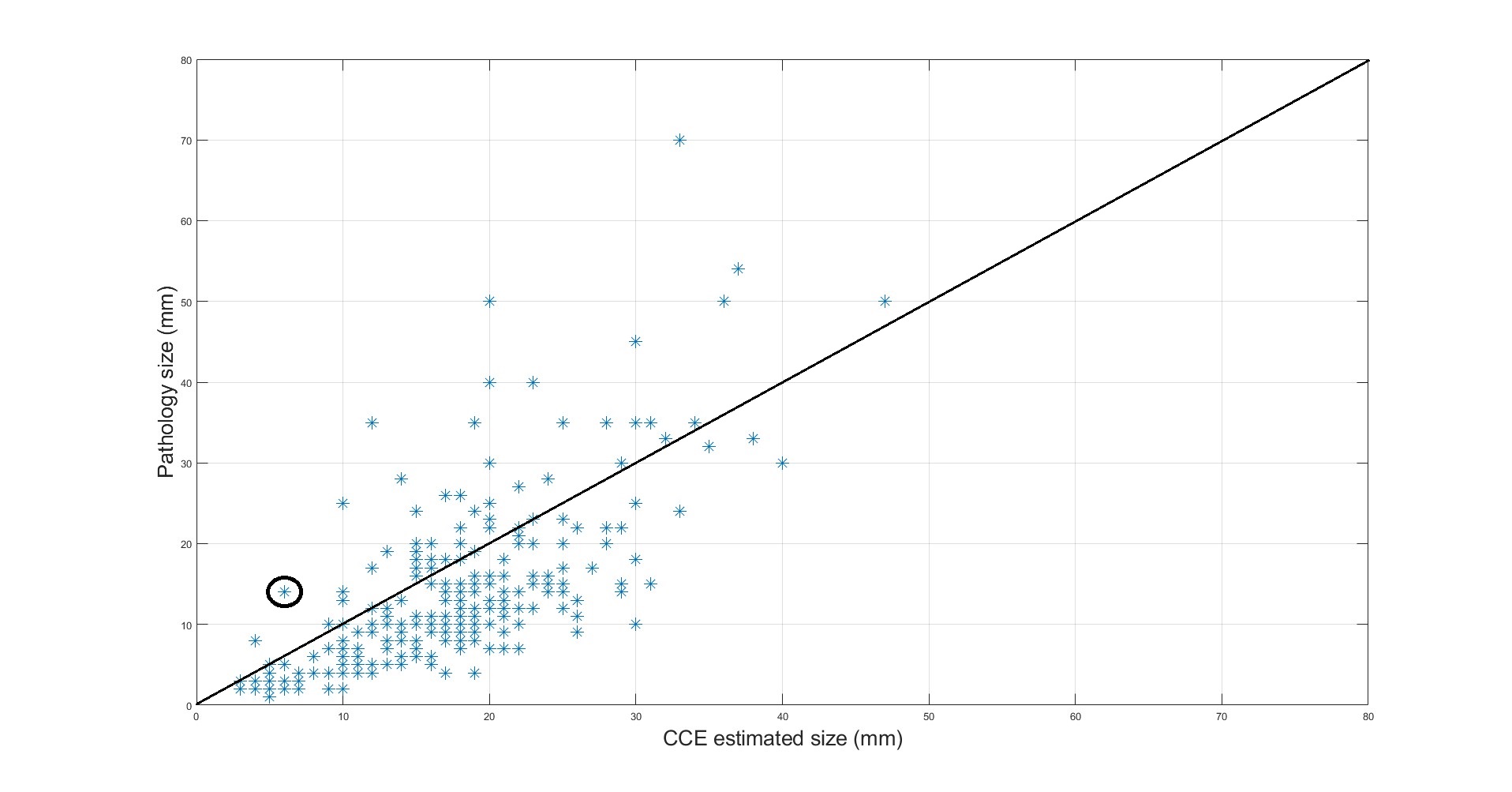}
\caption{Estimated size in CCE vs. histopathology (HP)}
\label{fig:CCE-Pathology}
\end{figure}

Setting the histopathology report as the golden standard, the next step in the size estimation algorithm was to map the estimated size from the segmented images to the outcomes from the pathology report, as shown in Figure \ref{fig:CCE-Pathology}. The number of certain matched polyps among the two databases (CCE vs. pathology) hitherto was $280$. Looking at figure \ref{fig:CCE-Pathology}, it is evident that the estimated polyp sizes from CCE are in general an overestimation of those retrieved from the pathology report. The best regression function estimator based on a fine Gaussian support vector machines (SVM) resulted in a root mean squared error (RMSE) of approximately $6mm$. It should be noted that despite the large error, a richer database of matched polyps seen by CCE and examined by pathologists will assist us in obtaining more accurate results. Enriching this database is currently an ongoing task for the completion of cfc2015 trial. 

\subsection*{Characterization}

Colorectal polyps are either neoplastic, or non-neoplastic. This classification, in addition to the size, histology, and the location of polyps (distal vs. proximal colon) dictate the management pathway of patient’s care, mode of actions, and therefore success of the treatment outcome. Our characterisation algorithm is a binary classifier, where those CCE images classified as important findings from the recognition network feature the input set, and two classes, namely neoplastic vs. non-neoplastic, forms the output. 

Comparable to the size estimation algorithm, the dataset features a total number of $144$ images, in which polyps observed during CCE investigations have been resected, and matched to those after polypectomy. Out of the $144$ images, $49$ corresponded to neoplastic lesions, and $95$ were non-neoplastic polyps. Given the small sample size, and class imbalance (ratio $2$ to $1$), we augmented the dataset fourfold, by horizontal and vertical random reflection, random scaling and random translation, along with random rotation, all picked from a continuous uniform distribution, without affecting the contents or the size of the images. It is worth mentioning that, only polyp's segment of an image should contribute with information on the pathology, and no feature from the background should influence the classification outcome. Despite this, our input dataset features the entire image frame, due to the fact that a big portion of images only contained small polyps. 

\begin{figure}[ht]
\centering
\includegraphics[width=14cm,height=10cm]{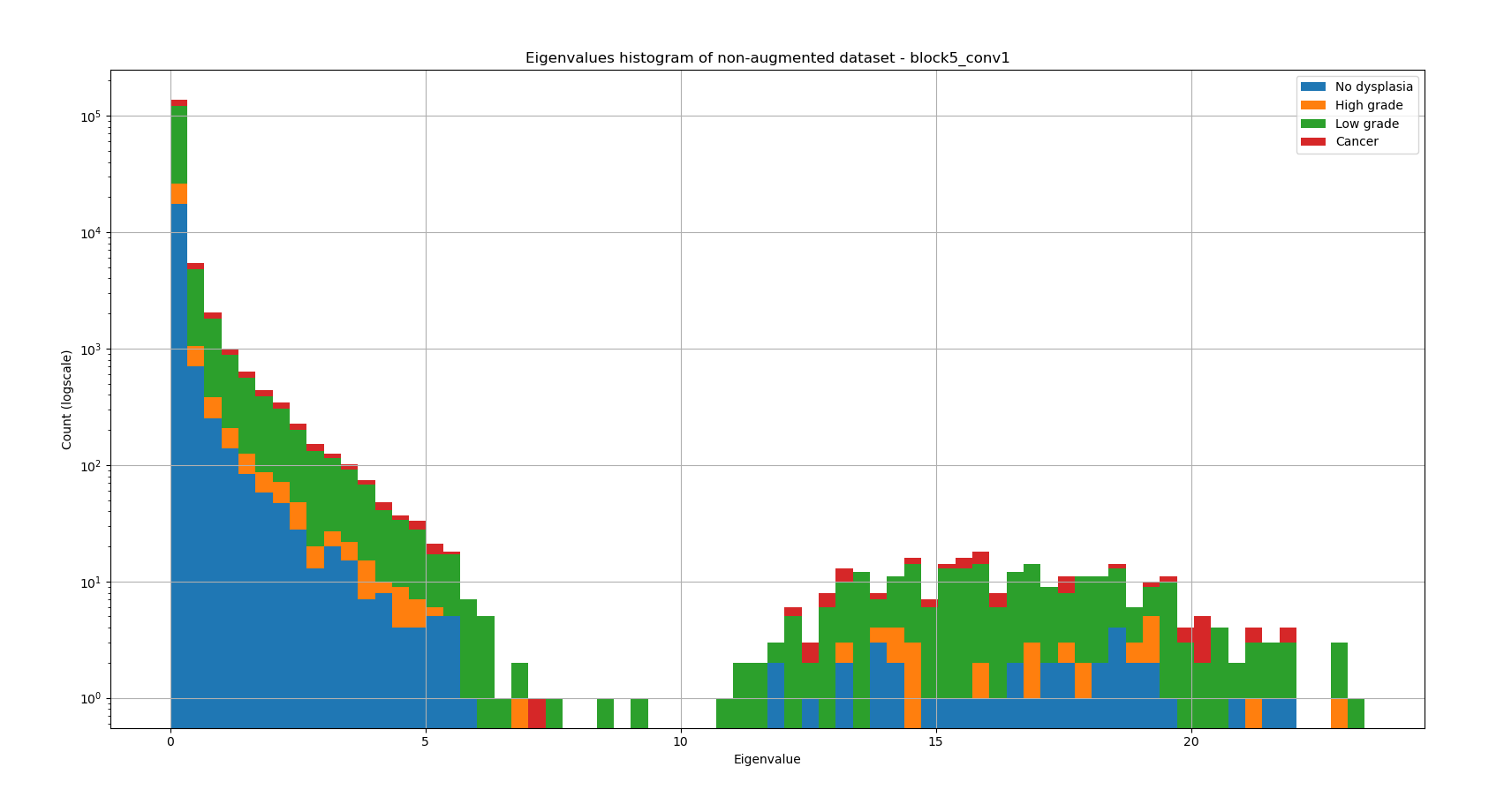}
\caption{Eigenvalue distribution of the VGG16 gram matrix.}
\label{fig:CCE-style}
\end{figure}

Following the same training settings as the recognition DNN, we deployed a VGG16 network as the backbone of the characterisation algorithm. Splitting the dataset to $70\%$ for training and validation, and the remaining $30\%$ for testing, resulted in a binary classifier with a sensitivity of $82\%$, a specificity of $80\%$, and an accuracy of $81\%$. To further improve the performance of the characterisation algorithm by incorporating information on the texture of polyps, we applied the neural style transfer (NST) strategy, which has demonstrated remarkable results for image stylization \cite{Gatys,Li}. Taking full advantage of the Gram matrices of the neural activation from different layers of a convolutional neural network (CNN), we obtained a representation of the style of the input image using a feature space originally designed to capture texture information. This feature space consisted of the correlations between different filter responses over the spatial extent of the feature maps, resulting in a stationary, multi scale representation of the input image that captured texture information but not the global arrangement. Keeping in mind that texture of a neoplastic polyp is different from a non-neoplastic one, this information can contribute to the evaluation of polyp's characteristics. We therefore complemented the outcome of VGG16 classification network with additional information drawn from the gram matrix of network's convolutional layers, i.e., the inner product between the vectorised feature maps of various layers. This was performed by looking into the distribution of both the bulk of the eigenvalues, and the largest eigenvalues of the gram matrix.
Our hypothesis was that depending on the pathology of the polyps (neoplastic vs. non-neoplastic), the supports of the distribution of the largest eigenvalues of the gram matrices would not overlap. Although this strategy has been successful in applications of anomaly detection and classification, the small sample size of only $144$ input images did not provide us with two distinct and separated supports for the distributions of the two classes. Similar pattern was observed when polyps were classified to "no dysplasia" vs. "low-grade and high-grade dysplasia" and cancer (Figure \ref{fig:CCE-style}). As a result, we only rely on the binary classification outcome from the trained VGG16 characterisation network, and we will revisit NST after the completion of cfc2015 trial, when the database is further enriched.  

\section*{Discussion and Conclusions}

Recently, the Danish National Institute of Health's (NIH) Technology Assessment Board has decided to not recommend AI as a decision support tool for colonoscopy when diagnosing neoplastic disease. The council stated that the rationale behind this decision is twofold: 1) Lack of sufficient evidence, where only a meta analysis of two randomized trials from the same authors sponsored by the manufacturing company is available, and 2) Outdated clinical guidelines with respect to the resection of all findings, and not accounting for the high number of extra diminutive, non-significant polyps. Currently, the national guidelines prescribe that all identified colorectal polyps, regardless of their size, morphology or pathology must be removed, and as AI is capable of detecting insignificant polyps as well as neoplastic ones, the treatment burden on both patients and the healthcare sector by deploying AI will, as a consequence, increase. While we are witnessing a surge of interest among hospitals to adopt AI-based solutions within their radiology departments, the decision of the Danish Treatment Council regarding AI-based colonoscopy draws a clear picture of the state of unreadiness of the healthcare sector to adopt AI as a decision support tool for the detection of gastrointestinal (GI) tract disorders. This discrepancy can be attributed to two factors: 1) Unlike in gastroenterology, in radiology a large body of publicly available databases can be accessed, which has lead to the development of robust models with high degree of generalizability, and 2) Instead of merely identifying the presence or absence of abnormalities, AI-based radiology reports convey complex diagnostic information with an appropriate degree of certainty. 

Clinical trials like cfc2015, and the ongoing efforts in both NHS Scotland (ScotCap) \cite{ScotCap} and NHS England \cite{NHS} to redesign outpatient gastroenterology services to enable early and effective screening by rolling out colon capsule endoscopy in the primary care to avoid unnecessary hospital appointments will certainly contribute to the availability of more data. Additionally, research studies like this work provides a set of AI-based solutions that are comparable to those deployed within radiology, bridging the existing gap between the state of AI deployment in gastroenterolgy vs. radiology. The overarching goal is that by completion of all the algorithms listed in Figure \ref{fig1}, we will be able to generate patient reports that convey complete diagnostic information, comparable to those within other clinical domains, where AI has a longer history of presence.

To reach this goal, each algorithm (Figure \ref{fig1}) must perform reliably. Proceeding in order, the recognition network has reached an unprecedented performance measure in terms of sensitivity, specificity and NPV. We therefore consider the recognition network to be ready for the external validation, where more than $3000$ previously unseen CCE videos from the ScotCap trial will be investigated. The network will, for each patient, report a set of candidate images containing polyps and other important findings. As part of future work, we will implement a range of explainability methods (X-AI), such as Pixel Rate-Distortion framework (pixel RDE) and CartoonX to improve our current gradient-based X-AI \cite{kutyniok}. This will allow us, for example, to better explain the partially incorrect classified regions as polyps, in the bottom image of Figure \ref{fig:stream}.

The first building block of the size estimation algorithm relies on the segmentation outcome from AID-U-Net(2,2). Despite showing a superior performance compared to the segmentation networks such as UNet and UNet++, where no specific pre-trained backbones were needed, AID-U-Net (2,2) obtained a comparable performance with UNetResNet (accuracy of $84\%$), where ResNet was deployed as the backbone architecture in the encoder network. It is worth mentioning that UNetResNet has a larger degree of complexity in terms of the number of learnable parameters compared to that of AID-U-Net. The second building block of the size estimation algorithm is a Gaussian SVM-based regression estimator that maps the estimated size of CCE findings to that of histopathology. Looking at Figure (\ref{fig:CCE-Pathology}), three points spark interest for further investigation: 1) the false negative marked by a circle, where CCE underestimated the size by a large margin, 2) the systematic overestimation of size by CCE compared to histopathology, and 3) the $6 mm$ size estimation error (RMSE) between CCE and histopathology.

The highlighted data point corresponds to a case where CCE underestimated the histopathology size by $8 mm$. This called for additional investigations of the entire CCE video. We found out that following the clinical guidelines, the polyp was matched (since it was the only polyp found in the right colon on both CCE and OC investigation). However, a thorough examination of the CCE video revealed that the polyp in CCE was next to the appendiceal orifice, and therefore certainly in the ceacum and not in the ascending colon. Therefore, the CCE polyp was incorrectly matched with the polyp described in OC. This specific case demonstrates the challenge of achieving adequate  interobserver agreement between CCE and OC, and intraobserver agreement within CCE evaluations. We need to  mention that for the size estimation algorithm, this data point was considered as outlier, and was therefore removed from the database of matched polyps. 

Overestimation of size in CCE compared to freshly retrieved (OC), and the formalin fixed polyps (histopathology) has been investigated by several authors \cite{blanes2018matching,Victoria1}, who demonstrated that substantial number of false positive CCE results are related to the assessment of polyps smaller than $6$ or $10$mm on OC and pathology, seen as larger than $6$ or $10$mm on CCE. While this discrepancy between CCE and OC can be attributed to the morphological assessment of the matched polyps, same rationale cannot justify the difference between CCE and pathology. Polyps seen by CCE appear more "pedunculated" compared to OC, and more "flat" when seen in OC compared to CCE. These differences in the morphology assessment may be due to the inflation of the colon during OC inspection, which may induce the polyps to appear more flattened in OC compared to CCE.

\begin{figure}[ht]
\centering
\begin{tikzpicture}
    \begin{axis}[
            colormap={bluewhite}{color=(white) rgb255=(90,96,191)},
            xlabel=Histopathology size (mm),
            xlabel style={yshift=-30pt},
            ylabel=CCE size (mm),
            ylabel style={yshift=20pt},
            xticklabels={HP $\leq6$,$7\leq$ HP $<10$, $10\leq$ HP $<20$ , HP$\geq 20$}, 
            xtick={0,...,4}, 
            xtick style={draw=none},
            yticklabels={CCE $\leq6$,$7\leq$ CCE $<10$, $10\leq$ CCE $<20$ , CCE$\geq 20$}, 
            ytick={0,...,4}, 
            ytick style={draw=none},
            enlargelimits=false,
            colorbar,
            xticklabel style={
              rotate=90
            },
            nodes near coords={\pgfmathprintnumber\pgfplotspointmeta},
            nodes near coords style={
                yshift=-7pt
            },
        ]
        \addplot[
            matrix plot,
            mesh/cols=4, 
            point meta=explicit,draw=gray
        ] table [meta=C] {
            x y C
            0 0 36
            1 0 1
            2 0 0
            3 0 0

            0 1 17
            1 1 1
            2 1 1
            3 1 0

            0 2 25
            1 2 26
            2 2 74
            3 2 14

            0 3 0
            1 3 5
            2 3 42
            3 3 38

        }; 
    \end{axis}
\end{tikzpicture}
\caption{Confusion matrix of estimated size: CCE vs. HP} \label{fig:M1}
\end{figure}
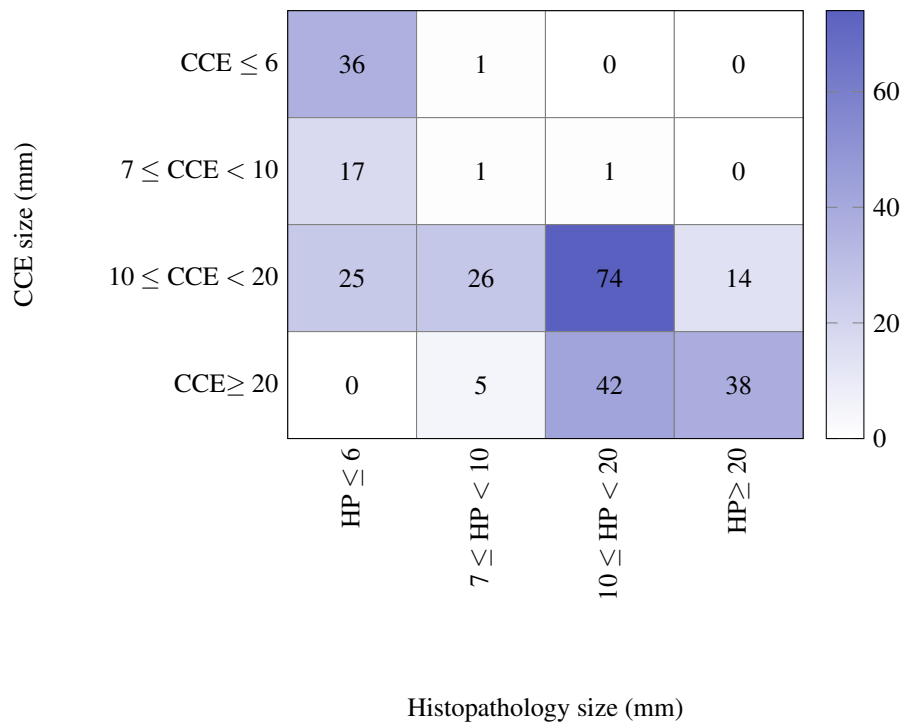

The $6$mm size estimation error (RMSE) between CCE and histopathology is partly a consequence of small number of data points, which will gradually improve once the database is further enriched. Looking at the confusion matrix in Figure \ref{fig:M1}, one can observe that the alteration of the regression-based size estimator to a classification-based one can improve the results. Splitting the estimated size to four classes, namely, $\leq 6mm$, $7mm \leq ... <10mm$, $10mm \leq ... < 20mm$, and $\geq 20mm$, in which the clinical decision for the next mode of action in patient's journey is based upon, can reduce the uncertainty in patient classification. \par

The two strategies we selected for the characterization algorithm, i.e., training a VGG16 network, and exploring texture information of the polyps using neural style transfer has been effective. Having said that, we expect that by further enriching the database in terms of both number of images, and inclusion of larger polyps, the performance of the algorithm will be improved. Inclusion of larger polyps and investigating texture information of only segmented portions (polyps) of the image, coupled with the information on the surrounding tissue will allow us to quantify the degree of vascularity. This will ease splitting the supports of the distribution of eigenvalues for neoplastic and non-neoplastic classes. 

While the external validation process of our developed algorithms using ScotCap data is in progress, continuous modification of these algorithms to improve their performance remains a priority. The challenging task of annotating the data, and reaching an interobserver and intraobserver agreement when evaluating CCE and OC investigations call for taking advantage of hybrid active learning approaches. Strategies such as Balance Exploration and Exploitation (BEE) and conformal prediction (CP), can actively query the teacher network for labels, resulting in a significantly lower number of required samples than in the normal supervised learning setting \cite{kutyniok}. Future work consists of completion of all the remaining algorithms, external validation using ScotCap data, and improving the performance of candidate algorithms. Our plan is to further adopt solutions from radiology workflow, such as PACS (picture archiving and communication system) and DICOM, and transfer CCE-generated non-DICOM images into cloud-based PACS for real-time analysis, report generation and sharing the outcomes with external medical professionals.

\section*{Data availability}

The data used in this article is part of the cfc2015 trial's outcome, which belongs to the "Odense University Hospital (OUH)". The request to access the data, or part of it should be made to Prof. Gunnar Baatrup.

\bibliography{main}

\section*{Acknowledgements}

This research is part of AICE project (number 101057400) funded by the European Union, and it is part-funded by the United Kingdom government. Views and opinions expressed are however those of the author(s) only and do not necessarily reflect those of the European Union or the European Commission. Neither the European Union nor the European Commission can be held responsible for them.

\begin{figure}[ht]
\centering
\includegraphics[width=8cm,height=2cm]{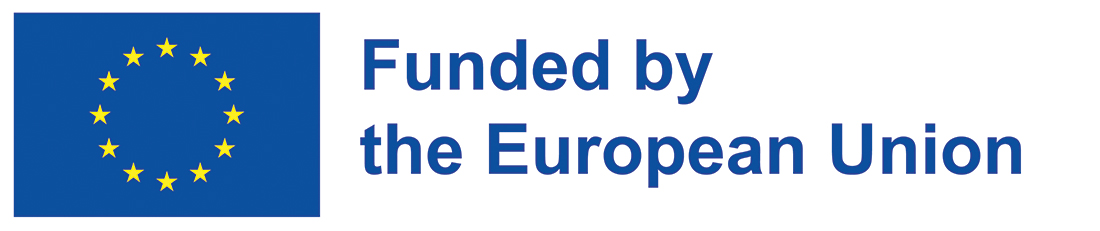}
\includegraphics[width=5cm,height=2cm]{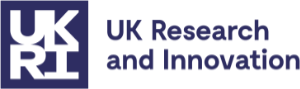}
\end{figure}

\section*{Author contributions statement}

E.S.N. and G.B. conceptualized the study. E.S.N., J.M.B, and E.P. developed the methods. B.S.O. and G.B. conducted the clinical trial and collected, and pre-processed the data. E.S.N., E.P. and V.B.V. analysed the results. E.S.N. and V.B.V. wrote the manuscript. All authors reviewed the manuscript. 

\section*{Additional information}

The authors declare no competing interests.

\end{document}